\documentclass[11pt,a4paper]{article}
\usepackage{jheppub}
\usepackage{cases}
\usepackage{graphicx}
\usepackage{bm}
\usepackage{epstopdf}
\usepackage{epsfig}
\usepackage{enumerate}
\usepackage{color,framed}
\usepackage{amsmath,amsthm,amssymb} 
\usepackage[normalem]{ulem}

\title{Collision of two kinks with inner structure}

\author[a]{Yuan Zhong,}
\author[b]{Xiao-Long Du,}
\author[a]{Zhou-Chao Jiang,}
\author[c]{Yu-Xiao Liu\footnote{Corresponding author.},}
\author[c]{Yong-Qiang Wang}

\affiliation[a]{School of Science, Xi'an Jiaotong University, Xi'an 710049, People's Republic of China}
\affiliation[b]{Carnegie Observatories, 813 Santa Barbara Street, Pasadena, CA 91101, USA}
\affiliation[c]{Institute of Theoretical Physics \& Research Center of Gravitation, Lanzhou University,
Lanzhou 730000, People' s Republic of China}
 \emailAdd{zhongy@mail.xjtu.edu.cn}
 \emailAdd{xdu@carnegiescience.edu}
 \emailAdd{j492667763@stu.xjtu.edu.cn}
 \emailAdd{liuyx@lzu.edu.cn}
 \emailAdd{yqwang@lzu.edu.cn}

\arxivnumber{1906.02920}

\abstract{In this work, we study kink collisions in a scalar field model with scalar-kinetic coupling. This model supports kink/antikink solutions with inner structure in the energy density. The collision of two such kinks is simulated by using the Fourier spectral method. We numerically calculate how the critical velocity and the widths of the first three two bounce windows vary with the model parameters. After that, we report some interesting collision results including two-bion escape final states, kink-bion-antikink intermediate states and kink or antikink intertwined final states. These results show that kinks with inner structure in the energy density have similar properties as those of the double kinks.}


\begin{document}
\maketitle  

\section{Introduction}
Kinks are topological defects in $1+1$ dimensional space-time, and have been applied in many areas of physics~\cite{Rajaraman1982,Vachaspati2006}. An important and interesting topic in the study of kinks is the interaction between kinks and antikinks. In integrable models, such as the sine-Gordon model, kink and antikink can pass each other after the collision with at most a phase shift~\cite{Das1989}. While in non-integrable models, the outcomes are more complex and sensitively depend on the initial velocities of kinks. Taking the $\phi^4$ model as an example, there exists a critical velocity $v_c\approx0.26$~\cite{Sugiyama1979}. When two kinks collide with a high initial velocity $v_0\geq v_c$, they simply bounce back after a collision; while when $v_0< v_c$, they form a bound state called bion (also known as oscillon)~\cite{Kudryavtsev1975}. Interestingly, in some intervals of velocity below $v_c$, instead of forming bion, kink and antikink finally escape after {\color{blue}a finite number of collisions}. These velocity intervals are called $m$-bounce windows ($m$BWs), if kinks collide $m$ times before bouncing back~\cite{Moshir1981,CampbellSchonfeldWingate1983}. All the bounce windows together form a fractal-like structure~\cite{AnninosOliveiraMatzner1991,GoodmanHaberman2007}. When generalized to higher dimensions, $\phi^4$ kinks can either describe a braneworld that we are living on~\cite{RubakovShaposhnikov1983}, or a bubble that we are living inside~\cite{KobzarevOkunVoloshin1975,Coleman1977,CallanColeman1977}. The collisions between both branes~\cite{KhouryOvrutSteinhardtTurok2001,KalloshKofmanLinde2001,Linde2001,TakamizuMaeda2004,TakamizuMaeda2006,GibbonsMaedaTakamizu2007,TakamizuKudohMaeda2007,Maeda2008,OmotaniSaffinLouko2011} and bubbles~\cite{HawkingMossStewart1982,GiblinHuiLimYang2010,ZhangPiao2010,AguirreJohnson2011,Kleban2011,WainwrightJohnsonPeirisAguirreEtAl2014,BondBradenMersini-Houghton2015,BradenBondMersini-Houghton2015,BradenBondMersini-Houghton2015a} have been extensively investigated in the literature. More works on interaction of $\phi^4$ kinks can be found in refs.~\cite{BelovaKudryavtsev1997,GoodmanHaberman2005}.

Recently, more and more {\color{blue} researchers} began to investigate kink interaction in other  {\color{blue} non-integrable} models, such as models with higher-order polynomial potentials~\cite{HoseinmardyRiazi2010,DoreyMershRomanczukiewiczShnir2011,Weigel2014,GaniKudryavtsevLizunova2014,GaniLenskyLizunova2015,Romanczukiewicz2017,BelendryasovaGani2017,SaxenaChristovKhare2018,GomesSimasNobregaAvelino2018,BelendryasovaGani2019,BazeiaMendoncaMenezesOliveira2019}, with various kinds of triangular potentials~\cite{PeyrardCampbell1983,GaniKudryavtsev1999,SimasGomesNobregaOliveira2016,SimasGomesNobrega2017,GaniMarjanehAskariBelendryasovaEtAl2018,BazeiaBelendryasovaGani2018,BazeiaGomesNobregaSimas2019a,BazeiaGomesNobregaSimas2019}, with generalized dynamics~\cite{GomesMenezesNobregaSimas2014}, and with multi-component scalar fields~\cite{AshcroftEtoHaberichterNittaEtAl2016,Alonso-Izquierdo2018b,Alonso-Izquierdo2018,Alonso-Izquierdo2018a,Alonso-IzquierdoBalseyroSebastianGonzalezLeon2018,Alonso-Izquierdo2019,Alonso-Izquierdo2019a}.

Some of these works renewed our understanding towards bounce windows. For example, it has been widely accepted that in order to form bounce windows, a kink should have a vibrational mode. It is the resonant energy transition between the vibrational mode and the translational mode that causes the formation of bounce windows. This mechanism was proposed by Campbell, Schonfeld and Wingate~\cite{CampbellSchonfeldWingate1983}, and has been successfully applied in many cases~\cite{CampbellPeyrardSodano1986,KivsharFeiVazquez1991,ZhangKivsharVazquez1992}. But some recent works have {\color{blue} shown} that even there is no vibrational mode around a single kink, bounce windows can still be formed~\cite{DoreyMershRomanczukiewiczShnir2011,GaniKudryavtsevLizunova2014,DoreyRomanczukiewicz2018}. On the other hand, more vibrational modes usually suppress the bounce windows~\cite{SimasGomesNobregaOliveira2016,BazeiaGomesNobregaSimas2019}. The development of the collective coordinate method~\cite{Weigel2014,TakyiWeigel2016,Weigel2018} and the discovery of the relation between bounce windows and the separatrix map~\cite{GoodmanHaberman2005} also help us to understand the bounce window phenomenon.

Some other studies found that in higher-order models like $\phi^8, \phi^{10}, \cdots$, due to the long-range interactions between kinks~\cite{MelloGonzalezGuerreroLopez-Atencio1998,GomesMenezesOliveira2012,KhareChristovSaxena2014,BelendryasovaGani2019,BazeiaMenezesMoreira2018,KhareSaxena2018,ChristovDeckerDemirkayaGaniEtAl2018}, the widely used superposition or production ansatz {\color{blue} is} problematic, and should be replaced by the so-called split-domain ansatz~\cite{ChristovDeckerDemirkayaGaniEtAl2019}. The force between long-range interacting kinks has been calculated recently~\cite{Manton2018,Manton2019}. There are also many other interesting topics on kink interaction, including multi-kink collision~\cite{SaadatmandDmitrievKevrekidis2015,MarjanehSaadatmandZhouDmitrievEtAl2017,MarjanehGaniSaadatmandDmitrievEtAl2017,MarjanehAskariSaadatmandDmitriev2018,EkomasovGumerovKudryavtsevDmitrievEtAl2018,GaniMarjanehSaadatmand2019}, boundary scattering~\cite{AntunesCopelandHindmarshLukas2004,ArthurDoreyParini2016,DoreyHalavanauMercerRomanczukiewiczEtAl2017,LimaSimasNobregaGomes2018}, negative radiation effect~\cite{ForgacsLukacsRomanczukiewicz2008,YamaletdinovRomanczukiewiczPershin2019}, creating kink-antikink pair by colliding particles or wave packages~\cite{DuttaSteerVachaspati2008,RomanczukiewiczShnir2010,DemidovLevkov2011a,DemidovLevkov2011,DemidovLevkov2015,AskariSaadatmandDmitrievJavidan2018}, spectral walls~\cite{AdamOlesRomanczukiewiczWereszczynski2019a,AdamOlesRomanczukiewiczWereszczynski2019}{\color{blue}. For} more related works, see ref.~\cite{RomanczukiewiczShnir2018a}.

In this paper, we will consider the collision of two kinks with inner structure in the energy density. In some models, especially models with generalized dynamics, as the parameter varies, the energy density of the kink might split from one peak to multi peaks~\cite{BazeiaLosanoMenezesOliveira2007,BazeiaLobaoMenezes2015a,ZhongGuoFuLiu2018}. When this happens, we say that the kink possesses an inner structure. Kinks with inner structure are similar to, but essentially different from double kinks~\cite{BazeiaMenezesMenezes2003}. Both structures have a local minimum at the center of the energy density function. But unlike the double kink case, where the local minimum at the center equals to zero, a kink with inner structure can have a nonzero local minimum at the center of the energy density function.

Collision between two double kinks was studied in many works, and some new interesting phenomena were found.
For example, two-bion escape final states were found in double sine-Gordon model~\cite{CampbellPeyrardSodano1986,GaniMarjanehAskariBelendryasovaEtAl2018}, and in sinh-deformed $\phi^4$ model~\cite{BazeiaBelendryasovaGani2018}. Unstable kink-bion-antikink intermediate states were found in refs.~\cite{MendoncaOliveira2015,MendoncaOliveira2015a}.

In this work, we will consider a model with coupling between the scalar field and its kinetic term. Such a generalized dynamics enables the kink to have rich and tunable inner structures. We will study the collision between a kink and an antikink of this model. Our model and corresponding static kink solution will be given in the next section. The numerical simulation of kink collision will be conducted in section \ref{secThree}. Finally, we will end this paper by a conclusion and outlook in section \ref{secFour}.

\section{The model, kink solution and its linear spectrum}
In our model, the scalar field $\phi$ is coupled {\color{blue} to} its kinetic term $X\equiv -\frac12\eta^{\mu\nu}\partial_\mu\phi\partial_\nu\phi$ via the following Lagrangian density:
\begin{eqnarray}
\mathcal{L}=G(\phi)X-V(\phi),
\end{eqnarray}
where $G(\phi)=1+\beta \phi^{2n}$. The parameter $\beta>0$ describes how much our model deviates from the canonical case ($\beta=0$), while the parameter $n=1,2,\cdots$ controls the number of local maxima in the energy density of the kink.

The equation of motion of our model is
\begin{eqnarray}
\label{eqEOM}
{G_\phi }({\partial _x}\phi {\partial _x}\phi  - {\partial _t}\phi {\partial _t}\phi ) + 2G(\partial _x^2\phi  - \partial _t^2\phi ) = 2{V_\phi },
\end{eqnarray}
{\color{blue} where the subscript $\phi$ denotes the derivative with respect to $\phi$.}
A static solution $\phi=\phi(x)$ can be obtained by solving the following equation:
\begin{eqnarray}
\label{eqEOMStatic}
\frac{1}{2}{G_\phi }{\partial _x}\phi {\partial _x}\phi  + G\partial _x^2\phi  = {V_\phi }.
\end{eqnarray}
A powerful method for constructing analytical static kink solutions is the superpotential method~\cite{BazeiaLosanoMenezes2008,ZhongLiu2014}, which begins with the assumption
\begin{eqnarray}
\label{super1}
{\partial _x}\phi=W(\phi).
\end{eqnarray}
By integrating the equation of motion \eqref{eqEOMStatic}, one can find a simple relation between the scalar potential and the superpotential $W$:
\begin{eqnarray}
\label{super2}
V = \frac{1}{2}G{W^2} + {V_0},
\end{eqnarray} where $V_0$ is an integral constant\textcolor[rgb]{0.00,0.00,1.00}{, which will be taken as zero}.

The superpotential formalism \eqref{super1}-\eqref{super2} makes it easy to find static kink solutions. For example, by taking
\begin{eqnarray}
\label{eqSuper}
W&=&k\phi_0\left[1-\left(\frac{\phi}{\phi_0}\right)^2\right],
\end{eqnarray}
one immediately obtains the $\phi^4$ type kink solution
\begin{eqnarray}
\label{solutionphi}
\phi&=&\phi_0\tanh(kx).
\end{eqnarray}
Here, $\phi_0$ represents the vacuum expectation value of $\phi(x)$, and $1/k$ the width of the kink. In this work, we will focus on the collision of this type of kink solution, and always take $k=\phi_0=1$ for simplicity. Other solutions will be considered in our future works.

\begin{figure}
\begin{center}
\includegraphics[width=0.7\textwidth]{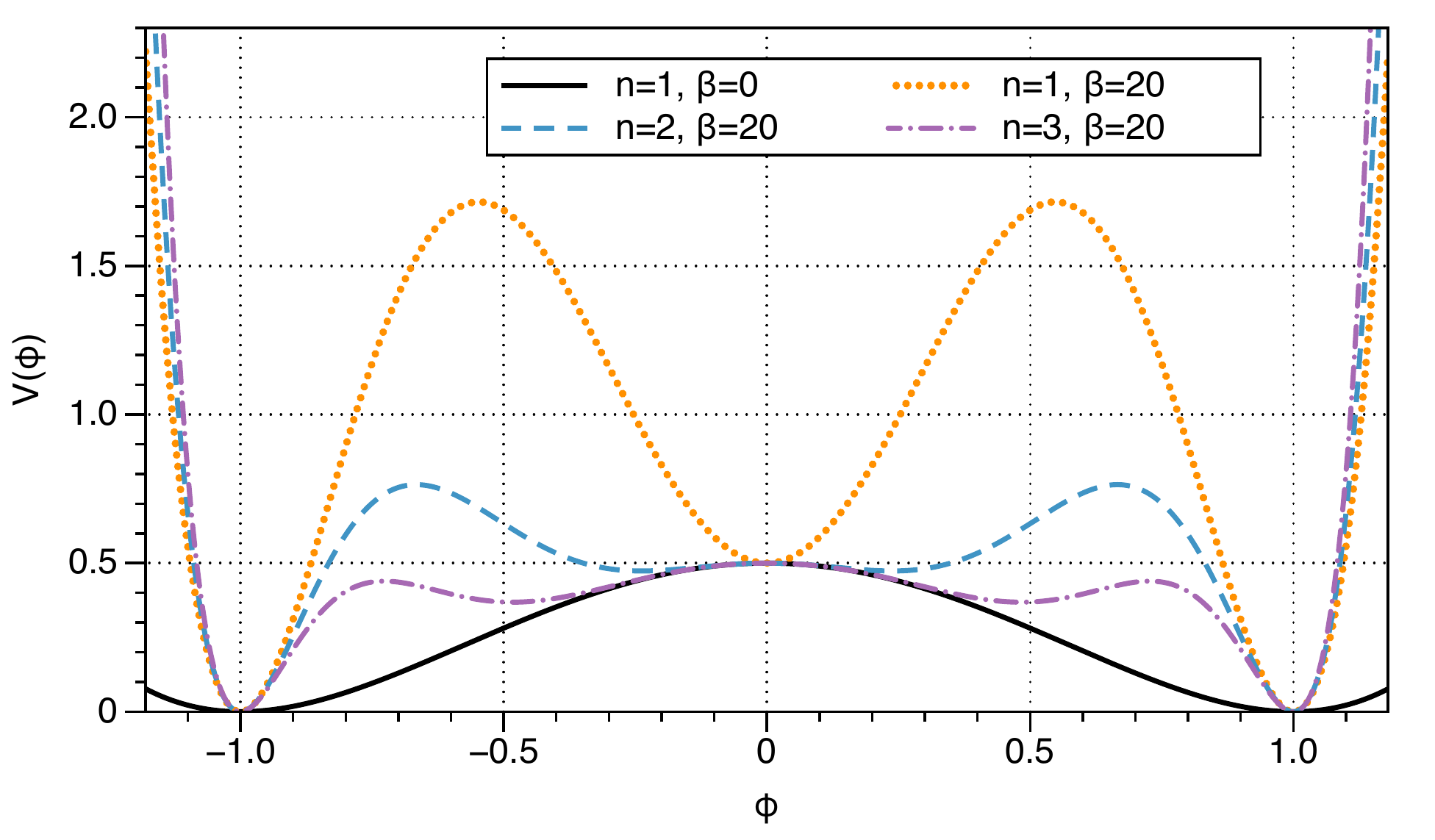}
\caption{The scalar potential $V(\phi)$.}
\label{fig_scalarPotential}
\end{center}
\end{figure}

\textcolor[rgb]{0.00,0.00,1.00}{The scalar potential takes the following form
\begin{eqnarray}
V=\frac{1}{2} \left(1-\phi ^2\right)^2 \left(\beta  \phi ^{2 n}+1\right),
\end{eqnarray} which is not the standard $\phi^4$ double-well potential when $\beta\neq 0$, see fig.~\ref{fig_scalarPotential}. }

\begin{figure}
\begin{center}
\includegraphics[width=0.8\textwidth]{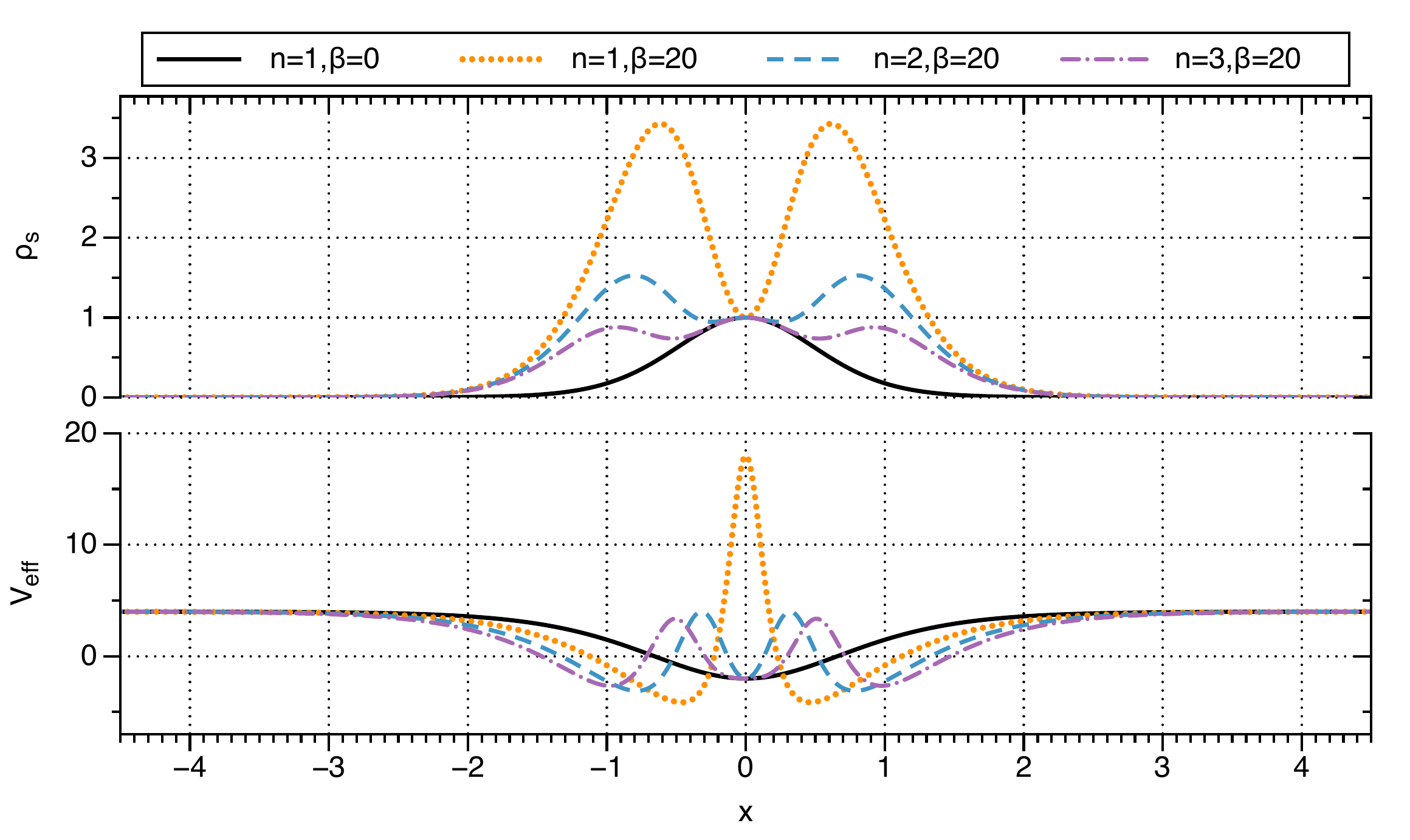}
\caption{The energy density $\rho(x)$ and the effective potential $V_{\textrm{eff}}$.}
\label{fig_Veff}
\end{center}
\end{figure}

The energy density of our model takes the form
\begin{eqnarray}
\rho &=&\frac{1}{2}G\left( \phi \right) \dot{\phi}^2+\frac{1}{2}G\left( \phi \right) \phi '^2+V\left( \phi \right).
\end{eqnarray}
 {\color{blue} In this work, we always use an overdot or a prime to denote the derivative with respect to time or space.}
For the static solution in eq.~\eqref{solutionphi}, the explicit expression of $\rho$ is
\begin{eqnarray}
\rho_s =\text{sech}^4(x) \left(\beta  \tanh ^{2 n}( x)+1\right),
\end{eqnarray}
whose shape is plotted in fig.~\ref{fig_Veff}. Obviously, $\rho_s$ splits if $\beta$ is large enough. Besides, the number of peaks of $\rho_s$ increases with $n$ for $n \leq 2$, and equals to three as $n>2$. Unlike the case of double kink, where $\rho_s(x=0)=0$, the solution here always satisfies $\rho_s(x=0)=1$.

Another important property of the static kink solution is its linear spectrum. Consider a small linear perturbation $\delta\phi(x,t)$ around the background kink solution $\phi(x)$. {\color{blue} Defining} $\psi\equiv \delta\phi\sqrt{G}$ and $\theta\equiv \phi'\sqrt{G}$, one may show that the equation for the perturbation to the first order is~\cite{ZhongLiu2014}
\begin{eqnarray}
\psi''-\frac{\theta''}{\theta}\psi-\partial_t^2\psi=0.
\end{eqnarray}
We can expand $\psi$ with the Fourier modes
\begin{eqnarray}
\psi=\sum_{a=0}^{\infty} f_a(x)e^{i\omega_a t},
\end{eqnarray}
where the mode functions satisfy a Schr\"odinger-like equation
\begin{eqnarray}
f_a''(x)-V_{\textrm{eff}}(x)f_a(x)=-\omega_a^2 f_a(x),
\end{eqnarray}
with the effective potential defined by
\begin{eqnarray}
V_{\textrm{eff}}&=&\frac{\theta''}{\theta}.
\end{eqnarray}
The explicit expression of $V_{\textrm{eff}}$ can be easily obtained after substituting {\color{blue} in} the kink solution. Here we only point out that when $k=\phi_0=1$, its asymptotic behavior is $V_{\textrm{eff}}(x\to\pm\infty)=4$, and its shape can be found in the lower panel of fig.~\ref{fig_Veff}.

\begin{figure}
\begin{center}
\includegraphics[width=0.9\textwidth]{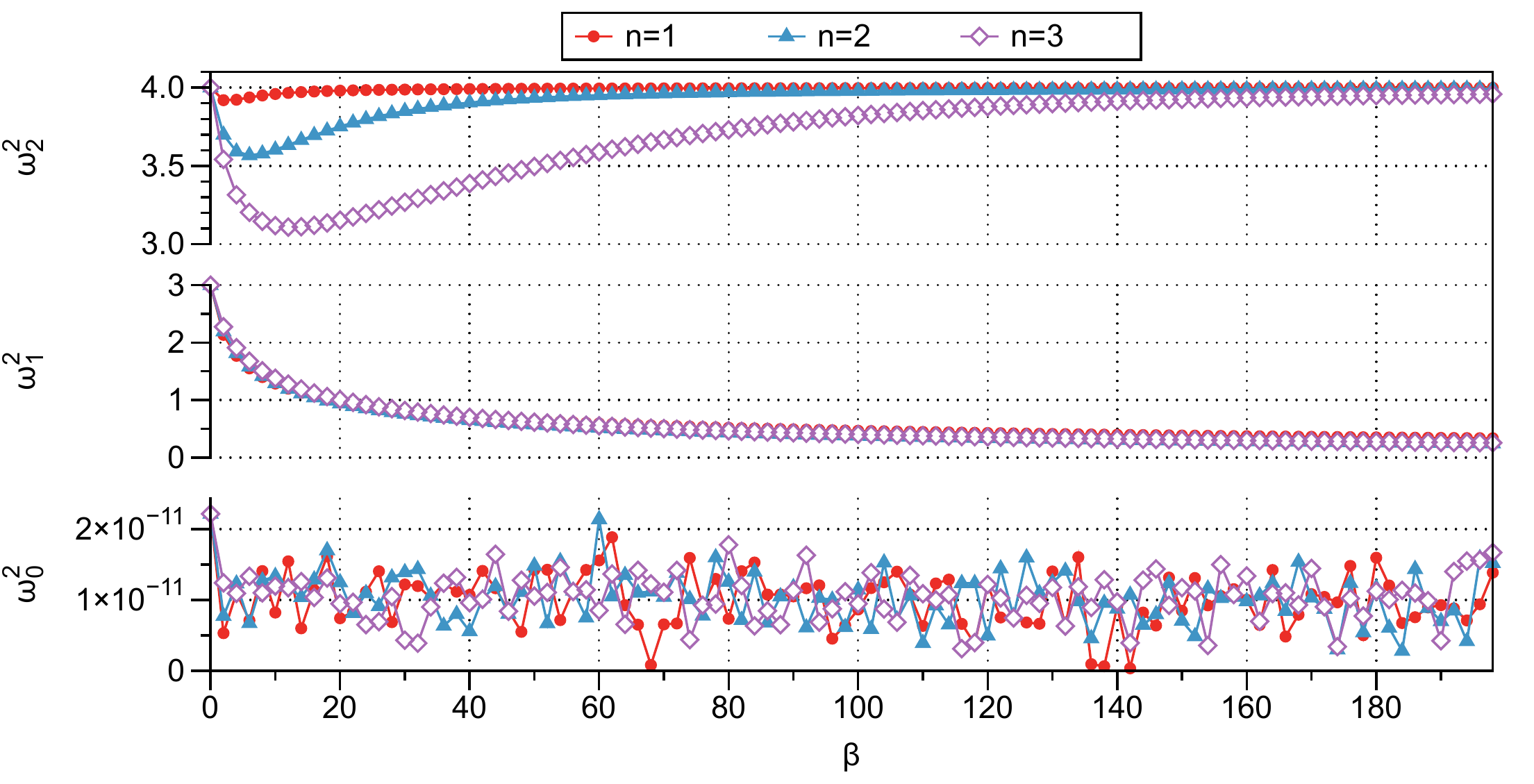}
\caption{The eigenvalues of the bound states $\omega_a^2$ {\color{blue} ($a=0,1,2$) with} the effective potential $V_{\textrm{eff}}(\beta,n;x)$, $n=1,2,3$ and $\beta\in[0:2:200]$.}
\label{fig_eingenvalues}
\end{center}
\end{figure}

The eigenvalues of the Schr\"odinger-like equation, $\omega_a^2$, can be calculated numerically. In fig.~\ref{fig_eingenvalues}, we plot the eigenvalues of all possible bound states for $n=1,2,3$ and $\beta\in[0:2:200]$. We find that in addition to the translational mode (the zero mode with frequency $\omega_0=0$), there are at most two vibrational modes, $\omega_1$ and $\omega_2$, in {\color{blue} the parameter ranges considered}. \textcolor[rgb]{0.00,0.00,1.00}{For small $\beta$, both $\omega_1$ and $\omega_2$ decrease as $\beta$ increases. But as  $\beta$ becomes larger, $\omega_1$ and $\omega_2$ behave differently: the former keeps {\color{blue} decreasing monotonically}, while the later increases after reaching a local minimum.}

\section{Kink-antikink collision}
\label{secThree}
In this section, we study the kink-antikink interaction.
Since we have no analytical multikink solution of our model, we will solve the dynamical equation numerically by taking the widely used superposition ansatz as the initial condition
\begin{subequations}
\label{initialCD}
\begin{eqnarray}
\phi(x,0)&=&\phi_K(-x_0,v_0; x, 0)+\phi_{\bar{K}}(x_0,-v_0;x, 0)-1,\\
\dot{\phi}(x,0)&=&\dot{\phi}_K(-x_0,v_0;x, 0)+\dot{\phi}_{\bar{K}}(x_0,-v_0;x, 0).
\end{eqnarray}
\end{subequations}
Here $\phi_K(x_0,v_0;x, t)=\tanh(\frac{x-x_0-v_0t}{\sqrt{1-v_0^2}})$ is a kink {\color{blue} initially located at $x_0$ and moving with an initial velocity of} $v_0<c=1$, and $\phi_{\bar{K}}(x_0,v_0;x, t)=-\phi_{K}(x_0,v_0;x,t)$ is the corresponding antikink solution. The solution of $\phi_K(x_0,v_0;x,t)$ is obtained by simply boosting the static kink in eq.~\eqref{solutionphi}. 

For simplicity, we will take periodical boundary condition, and solve the dynamical equation by using the Fourier spectral method. In this method, \textcolor[rgb]{0.00,0.00,1.00}{
$N$ evenly spaced grid points, or collocation points, are chosen on a finite truncated space domain. The solution of the scalar field is approximated by a truncated Fourier series
\begin{eqnarray}
\label{phi_approx}
\phi(x,t)\approx \phi^N (x,t)=\sum_{a=1}^{N} f(k_a,t)  e^{i k_a x},
\end{eqnarray}
where the coefficients $f(k_a,t)$ can be determined by requiring that $\phi^N=\phi$ at all collocation points. The $j$-th order spatial derivative of the solution is then approximated by differentiating $\phi^N$:
\begin{eqnarray}
\label{phi_approx_dn}
\partial_x^j \phi^N (x_b,t)=\sum_{a=1}^{N} (i k_a)^j f(k_a,t)  e^{i k_a x_b}\equiv \sum_{c=1}^{N} D_{bc}^j \phi^N(x_c,t) .
\end{eqnarray}
Here $D^j$ is a constant matrix called the derivative matrix and we have used the fact that $f(k_a,t)$ is a linear combination of $\phi$ at collocation points\footnote{The specific form of the derivative matrix can be found in \cite{Trefethen2001}.}.
}

As a result, the original partial differential equation (PDE) \eqref{eqEOM} will be transformed into a system of second-order-in-time ordinary differential equations (ODEs), which can be easily solved by using the $\text{ode45}$ solver of Matlab (see also refs.~\cite{Trefethen2001,ChristovDeckerDemirkayaGaniEtAl2018,ChristovDeckerDemirkayaGaniEtAl2019}). \textcolor[rgb]{0.00,0.00,1.00}{The numerical precision of this method is determined by two factors: the spatial step size, which changes with the number of the collocation points $N$; and the time step size, which will be automatically determined by the $\text{ode45}$ solver in accordance with its step changing algorithm. To improve the precision, one could add more collocation points and tune the relative and absolute tolerance options of the ode45 solver\footnote{In Matlab, the default value of the relative and absolute tolerance of ode45 solver are RelTol=$10^{-3}$ and AbsTol=$10^{-6}$. Usually, for a fixed number of collocation points $N$, more precise solutions can be obtained by taking smaller RelTol and AbsTol.}.}

\begin{figure}
\begin{center}
\includegraphics[width=0.8\textwidth]{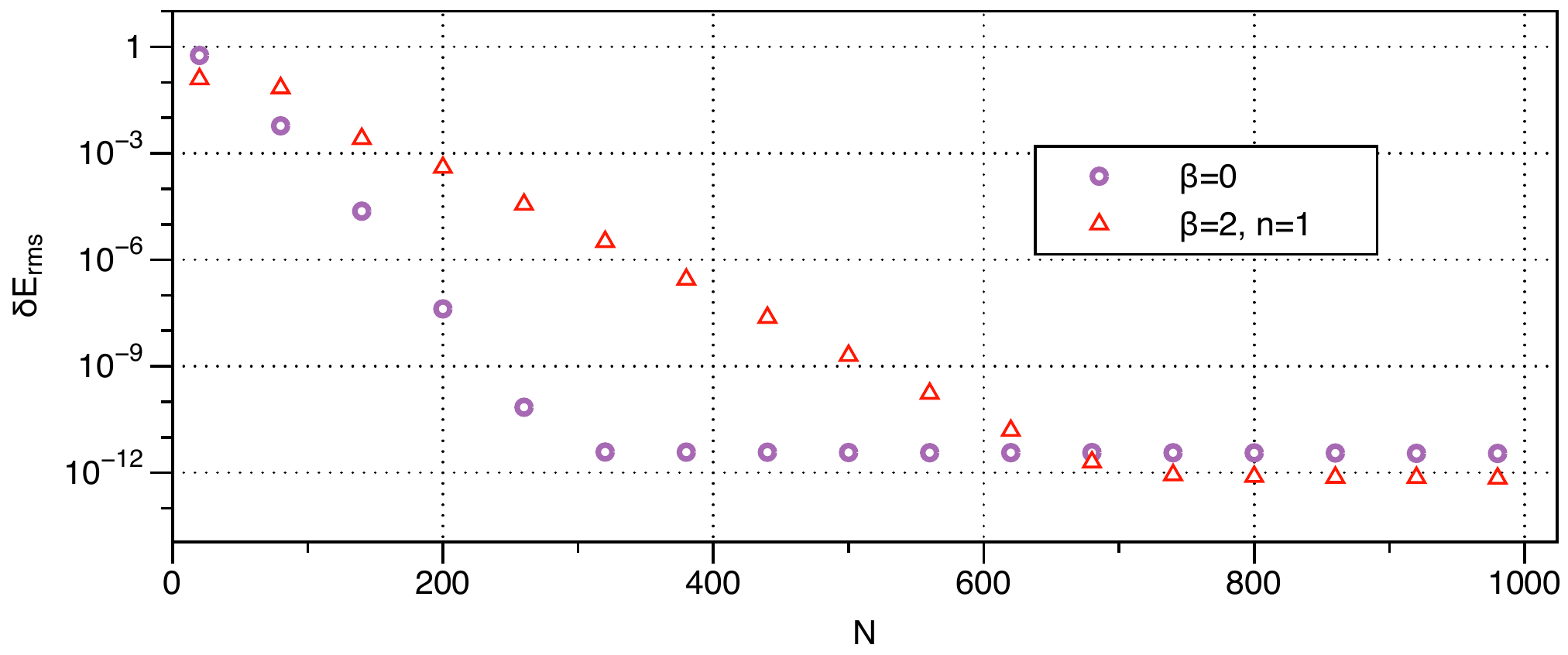}
\caption{ $\delta E_{\textrm{rms}}$ as a function of the number of collocation points $N$. The tolerance option of the ode45 solver is set as RelTol$=10^{-11}$ and AbsTol$=10^{-13}$.}
\label{fig_dE_N}
\end{center}
\end{figure}

\textcolor[rgb]{0.00,0.00,1.00}{To check the viability of our numerical results, we test the conservation of the total energy as the time evolution proceeded. For a single kink/anti-kink moving with velocity $v_0$, the energy is
\begin{eqnarray}
E(v_0)=\frac{\int_{-\infty}^{+\infty}\rho_s dx }{\sqrt{1-v_0^2}}=\left(\frac{4}{3}+\frac{4 \beta }{4 n (n+2)+3}\right)/\sqrt{1-v_0^2}.
\end{eqnarray}
Then, for a system of a pair of widely separated kink and anti-kink, the exact total energy is $E_{\textrm{exact}}=2E(v_0)$, which should be conserved all the time. We can compare our numerical solution $E_{\textrm{num}}(t)$ with the exact one by defining the relative error of total energy at time $t$ as
\begin{eqnarray}
\delta E(t)\equiv \frac{E_{\textrm{exact}}-E_{\textrm{num}}(t) }{E_\textrm{exact}}.
\end{eqnarray}
Because $\delta E(t)$ changes slightly with $t$, it would be more convenient to use the root mean square error $\delta E_{\textrm{rms}}$ to estimate the long-term behavior of energy conservation.  As examples, we consider two cases with $\beta=0$ and $\beta=2, n=1$, respectively. The former case is just the standard $\phi^4$ model. For both cases we take $x_0=10, v_0=0.18$ and conduct simulations within the spatial domain $x\in [-40, 40]$ for $t\in[0, 100]$. As can be seen from fig.~\ref{fig_dE_N}, $\delta E_{\textrm{rms}}$ converges exponentially as the collocation points $N$ increases, until it reaches the relative tolerance of the ode45 solver.
 }

\subsection{Critical velocity and two bounce windows}
Our model reduces to the well studied $\phi^4$ model when $\beta=0$. Therefore, it is interesting to see how a nonzero $\beta$ would change the well-known properties of the $\phi^4$ model.
In this section, we will consider the impacts of $\beta$ and $n$ on the value of critical velocity $v_c$ and on the widths of the two bounce windows.

\begin{figure}
\begin{center}
\includegraphics[width=0.8\textwidth]{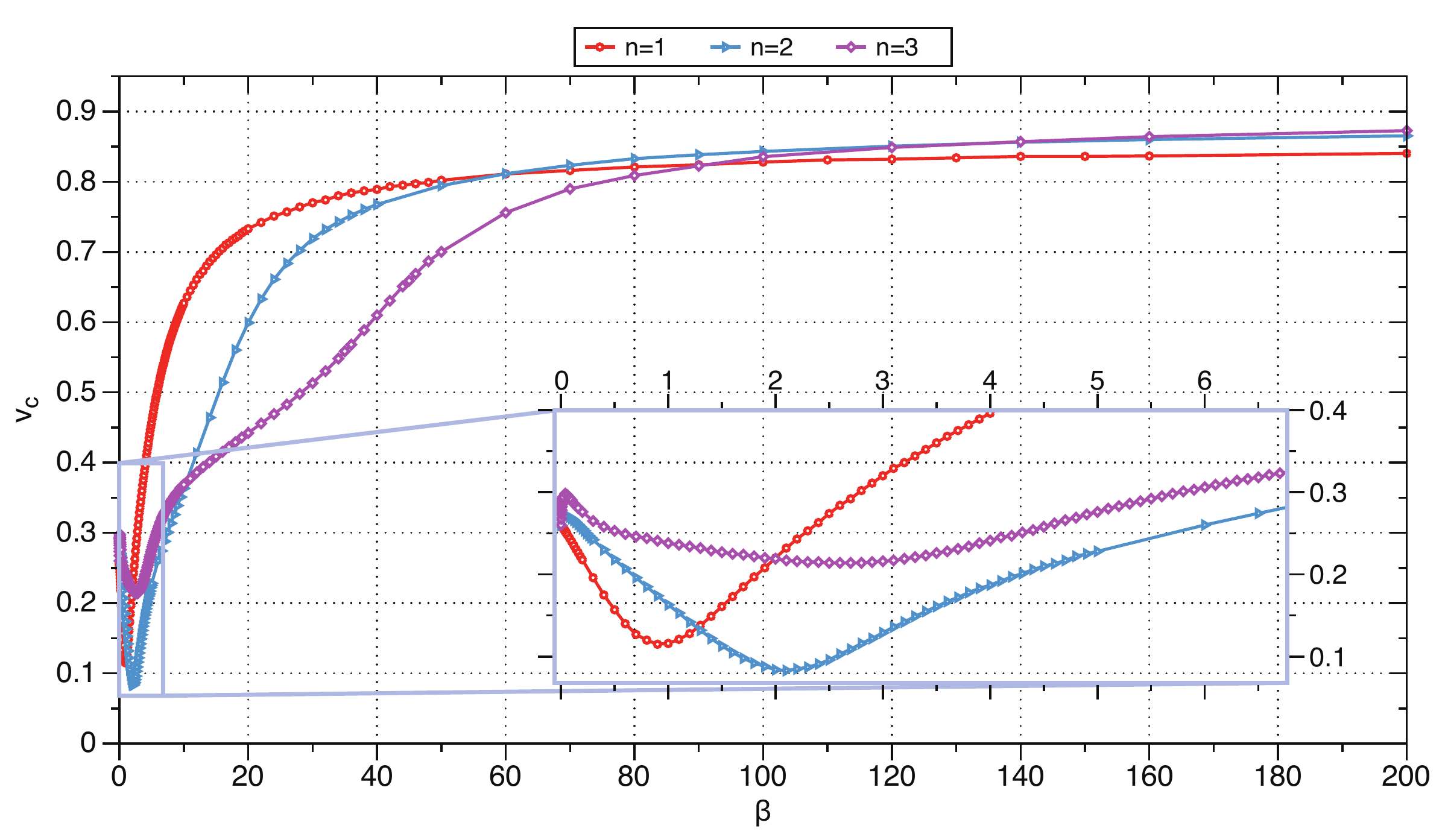}
\caption{The critical velocity as a function of $\beta$ in the case of $n=1,2,3$.}
\label{fig_vc}
\end{center}
\end{figure}

 In fig.~\ref{fig_vc}, we plot the critical velocity as a function of the parameter $\beta$ for cases with $n=1, 2, 3$. For different values of $n$, the global behavior of $v_c$ is similar: it has a global minimum around $\beta_{\text{min}}\approx n$, and increases monotonically as $\beta> \beta_{\text{min}}$. When $\beta=200$, the critical velocity increases to about 0.85 for $n=1, 2, 3$. It is also interesting to note that for $n=2, 3$, $v_c$ has a local maximum around $\beta=0.04$, see table \ref{tableMaxMinvc}.
\begin{table}[h]
\begin{center}
\begin{tabular}{|c|c|c|c|}
  \hline
   & $n=1$ & $n=2$ & $n=3$ \\
   \hline
  $(\beta, v_c)_{\textrm{max}}$ & --- & (0.04, 0.2711) & (0.04, 0.2982) \\
  \hline
  $(\beta, v_c)_{\textrm{min}}$& (0.9, 0.115) & (2.1, 0.083) & (2.8, 0.214) \\
  \hline
\end{tabular}
\end{center}
\caption{The local maxima and minima of $v_c$ for $n=1,2,3$, and $\beta\in[0, 200]$.}
\label{tableMaxMinvc}
\end{table}

\begin{figure}
\begin{center}
\includegraphics[width=0.7\textwidth]{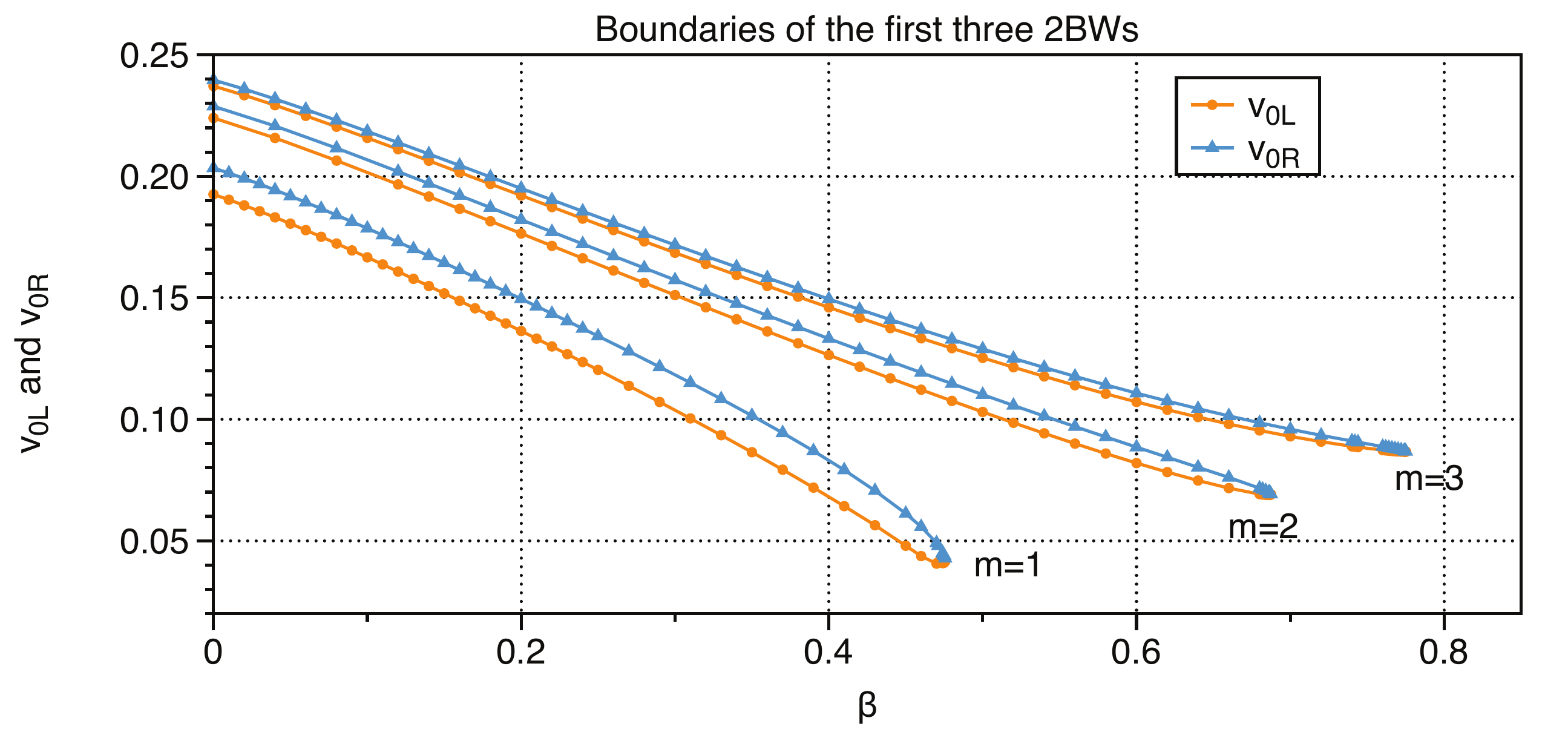}
\includegraphics[width=0.7\textwidth]{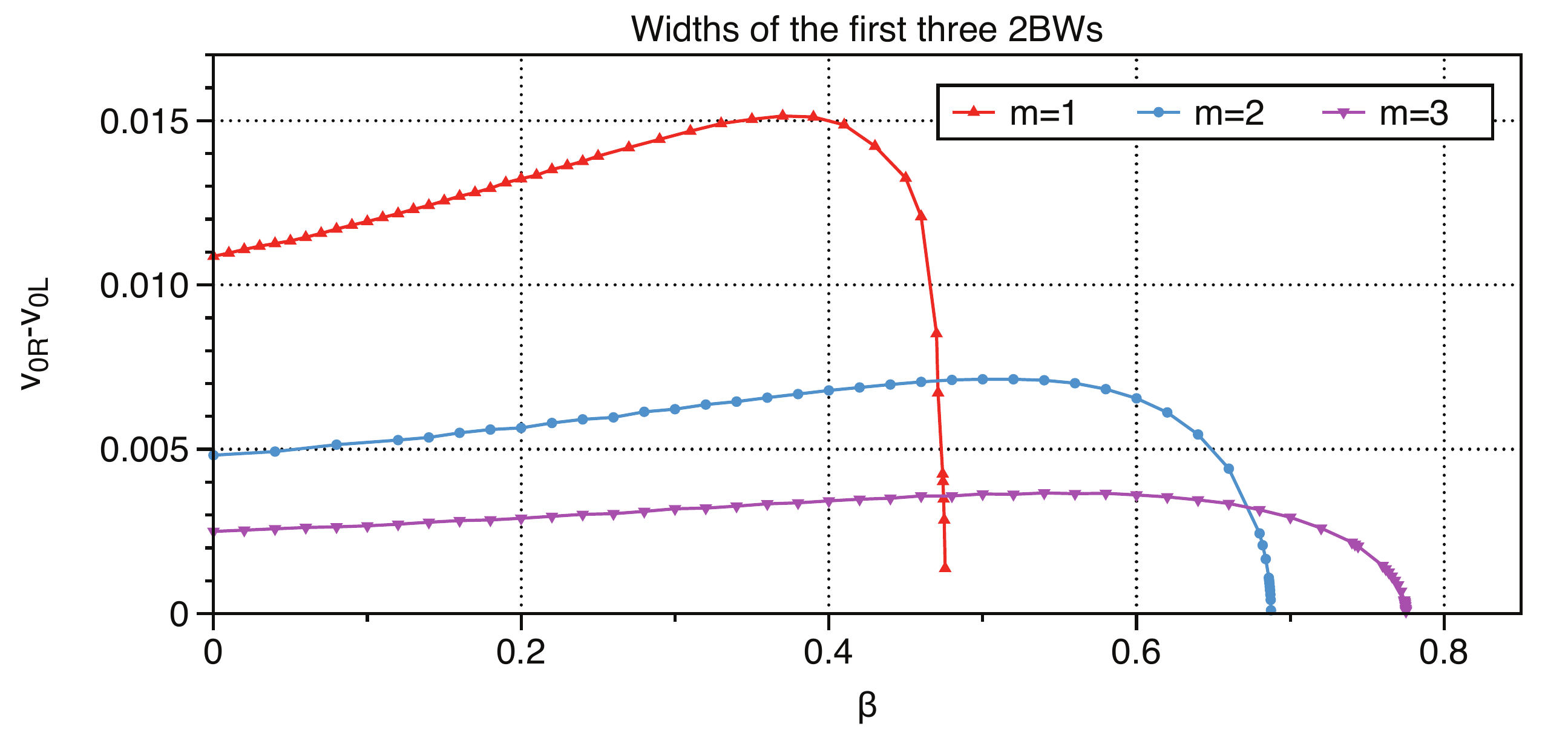}
\caption{The boundaries (upper panel) and the widths (lower panel) of the first three 2BWs (labeled by $m=1, 2, 3$) for $n=1$ and $\beta\in[0, 0.8]$. Here the boundaries and the widths are calculated at $t=200$. $v_{0L}$ and $v_{0R}$ are the left and right boundaries of the two bounce windows.}
\label{fig_WidthOf2BW}
\end{center}
\end{figure}

The model parameters also have impacts on the widths of the two bounce windows (2BWs). Figure~\ref{fig_WidthOf2BW} shows how the boundaries (the upper panel) and the widths (the lower panel) of the first three 2BWs (labeled by $m=1,2,3$, respectively) vary with $\beta$ in the case with $n=1$.  As can be seen from the figure, when $\beta$ increases the 2BWs expand {\color{blue} slightly} at the beginning, then {\color{blue} shrink rapidly}, and finally close when $\beta$ is large enough. For $n=1$, the first three 2BWs close at $\beta\approx 0.4756, 0.6874, 0.77515$, respectively.

\begin{figure}
\begin{center}
\includegraphics[width=1\textwidth]{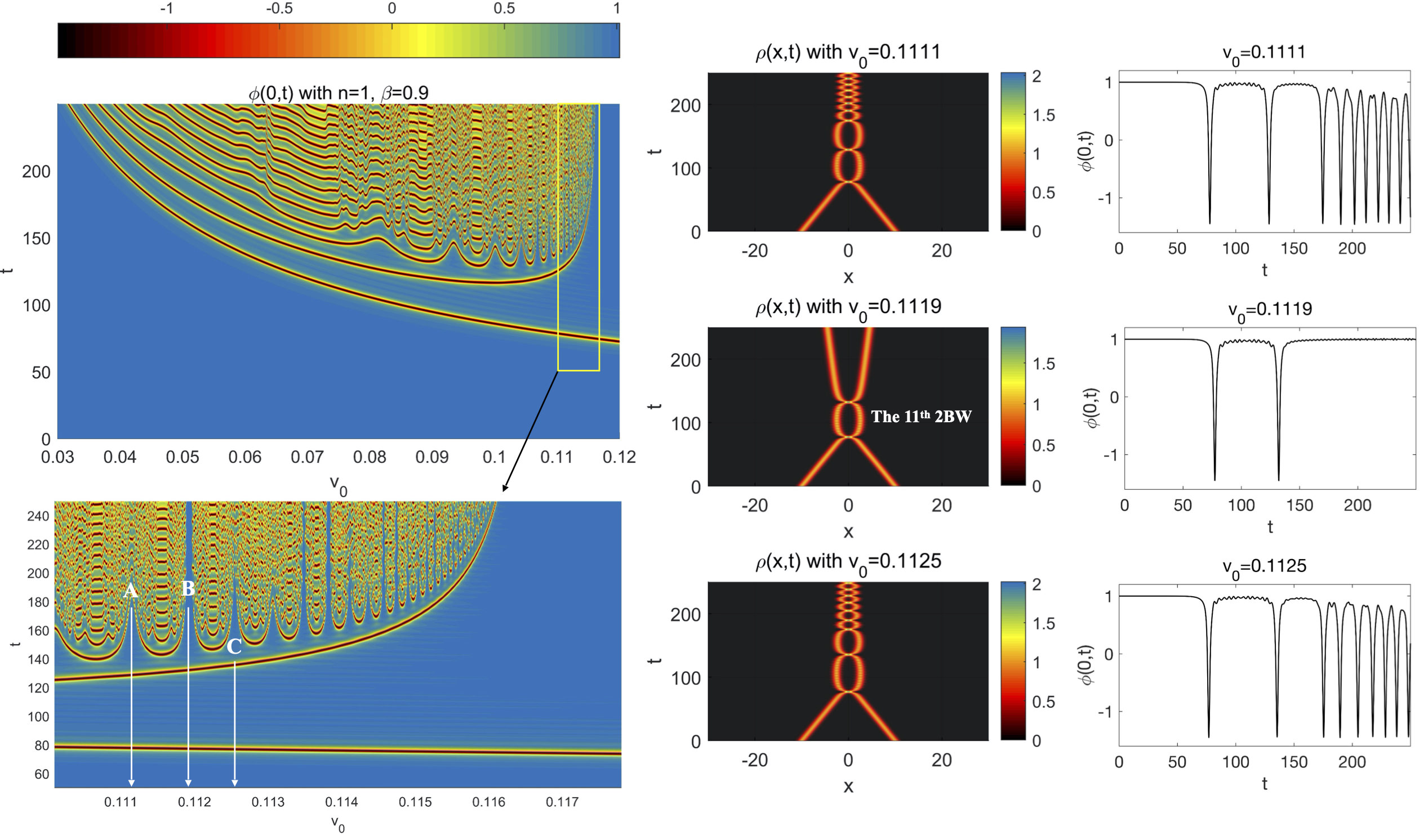}
\caption{Plot of $\phi(x=0,t)$ as a function of $v_0$ for $n=1$ and $\beta=0.9$. Under this parameter setting, the critical velocity is $v_c\approx 0.116$, and the widest two bounce window lies around $v_0\approx0.112$, which is the eleventh two bounce window. Since some higher-order two bounce window (the thirteenth, for instance) are already closed, we can conclude that the two bounce windows do not close order by order as $\beta$ increases.}
\label{fig_phi0t_beta09}
\end{center}
\end{figure}

One may guess that, as $\beta$ increases further, the fourth, fifth $\cdots$ 2BWs will close order by order. To {\color{blue} test} this, let us consider the case with $n=1, \beta=0.9$. In order to get a global view on the collision results, {\color{blue} we} consider $\phi(x=0)$ as a function of $t$ and $v_0$. When $v_0$ is fixed, the function $\phi(0,t)\equiv\phi(x=0,t)$ traces out a curve, which has many local minima with each {\color{blue} corresponding} to a collision of kinks (some examples can be found in the third column of fig.~\ref{fig_phi0t_beta09}). While, if $v_0$ varies, the local minima form a complex pattern, from which we can easily see the distribution of $m$BWs and bions. In $\phi(0,t)$ figure, an $m$BW is simply an interval of $v_0$ with $m$ dark lines.

In the first column of fig.~\ref{fig_phi0t_beta09}, we plotted $\phi(x=0)$ \textcolor[rgb]{0.00,0.00,1.00}{in the range $t\in[0, 240]$} and $v_0\in[0.03:0.0001:0.12]$. The numerical calculation is conducted by setting  the initial separation of the kinks as $2x_0=20$, and \textcolor[rgb]{0.00,0.00,1.00}{ taking 400 collocation points in the domain  $x\in[-50, 50]$. The tolerance option of ode45 solver is set as RelTol=$10^{-9}$ and AbsTol=$10^{-10}$, which {\color{blue} ensures} that $\delta E_{\textrm{rms}} \sim 10^{-9}-10^{-10}$.}  From fig.~\ref{fig_phi0t_beta09}, we can roughly estimate the value of critical velocity ($v_c\approx0.116$) and figure out the locations of 2BWs. For example, by magnifying the interval $v_0\in[0.110, 0.117]$ we find a clear 2BW around $v_0=0.1119$.

In the middle column of fig.~\ref{fig_phi0t_beta09}, we plot the energy densities correspond to three different initial velocities: A $(v_0=0.1111)$, B $(v_0=0.1119)$ and C $(v_0=0.1125)$. In the cases A and C, kinks collide many times at $x=0$, which indicates the forming of bions. While in the case B, kinks only collide twice before escaping, and is a two bounce collision. From the $\phi(0,t)$ figure of B, we clearly see that there are twelve local maxima between the two collisions, so B belongs to the $11^{\textrm{th}}$ 2BW. Point C locates at the center of the $12^{\textrm{th}}$ 2BW, which has been closed. So we can conclude that as $\beta$ increases further, the 2BWs do not closed order by order.  {\color{blue} We check these numerical results by taking $800$ collocation points.}

\subsection{Interesting intermediate and final states}
\label{secCollision}
In this section, we report some of the interesting phenomena in cases with large $\beta$. For large $\beta$, kinks can have rich structures in their energy densities. As we will see in this section, the collision of two kinks with inner structure can generate some interesting intermediate and final states.

One of the interesting phenomena is the escape of two bions, which has been found and discussed in many models such as the double Sine-Gordon model~\cite{CampbellPeyrardSodano1986,GaniMarjanehAskariBelendryasovaEtAl2018,GaniMarjanehSaadatmand2019}, the sinh-deformed $\phi^4$ model~\cite{BazeiaBelendryasovaGani2018} and other models with double kinks~\cite{MendoncaOliveira2015,MendoncaOliveira2015a,MendoncaDeOliveira2019}. In previous works, two-bion final states are usually generated by colliding a pair of double kinks. In this work, we find that when noncanonical dynamics is considered, it is also possible to generate two-bion escape final state from a kink-antikink initial state.

\begin{figure}
\begin{center}
\includegraphics[width=1\textwidth]{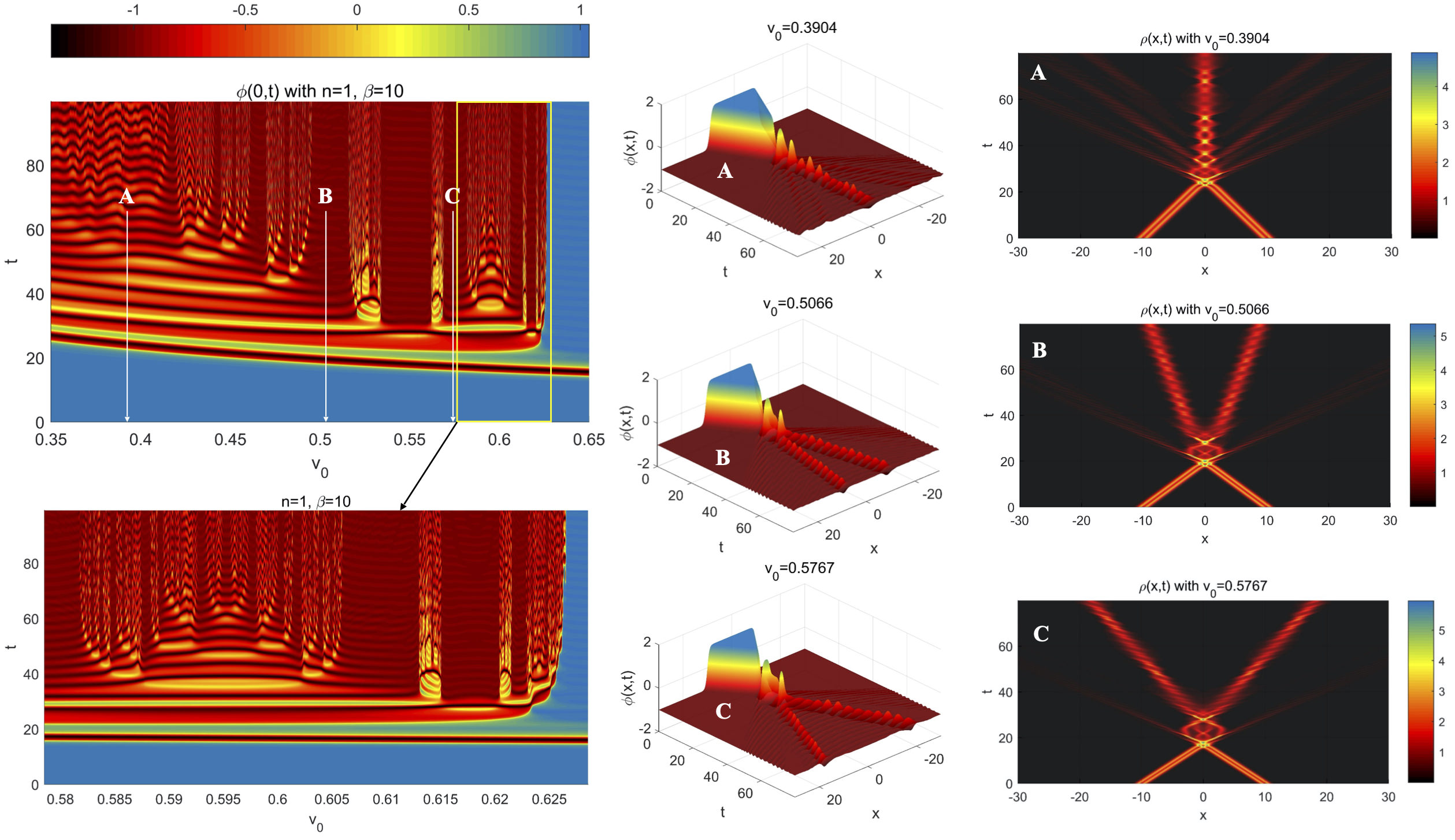}
\caption{For $n=1, \beta=10$, $v_0\in[0.35:0.00005:0.65]$  we find two wider two-bion escape windows in the vicinity of $v_0=0.5$ and $v_0=0.55$, respectively. We also plot the scalar configurations and the energy densities corresponding to three initial velocities denoted by A, B and C. Point A ($v_0=0.3904$) corresponds to a bion oscillates at $x=0$,  while B ($v_0=0.5066$) and C ($v_0=0.5767$) are examples of two escaping bions. In the present set of parameters, the energy density has two peaks (see the third column), and therefore the kinks of our model have similar properties as the double kink found in double sine-Gordon model. }
\label{fig_n1_beta_10}
\end{center}
\end{figure}

\begin{figure}
\begin{center}
\includegraphics[width=1\textwidth]{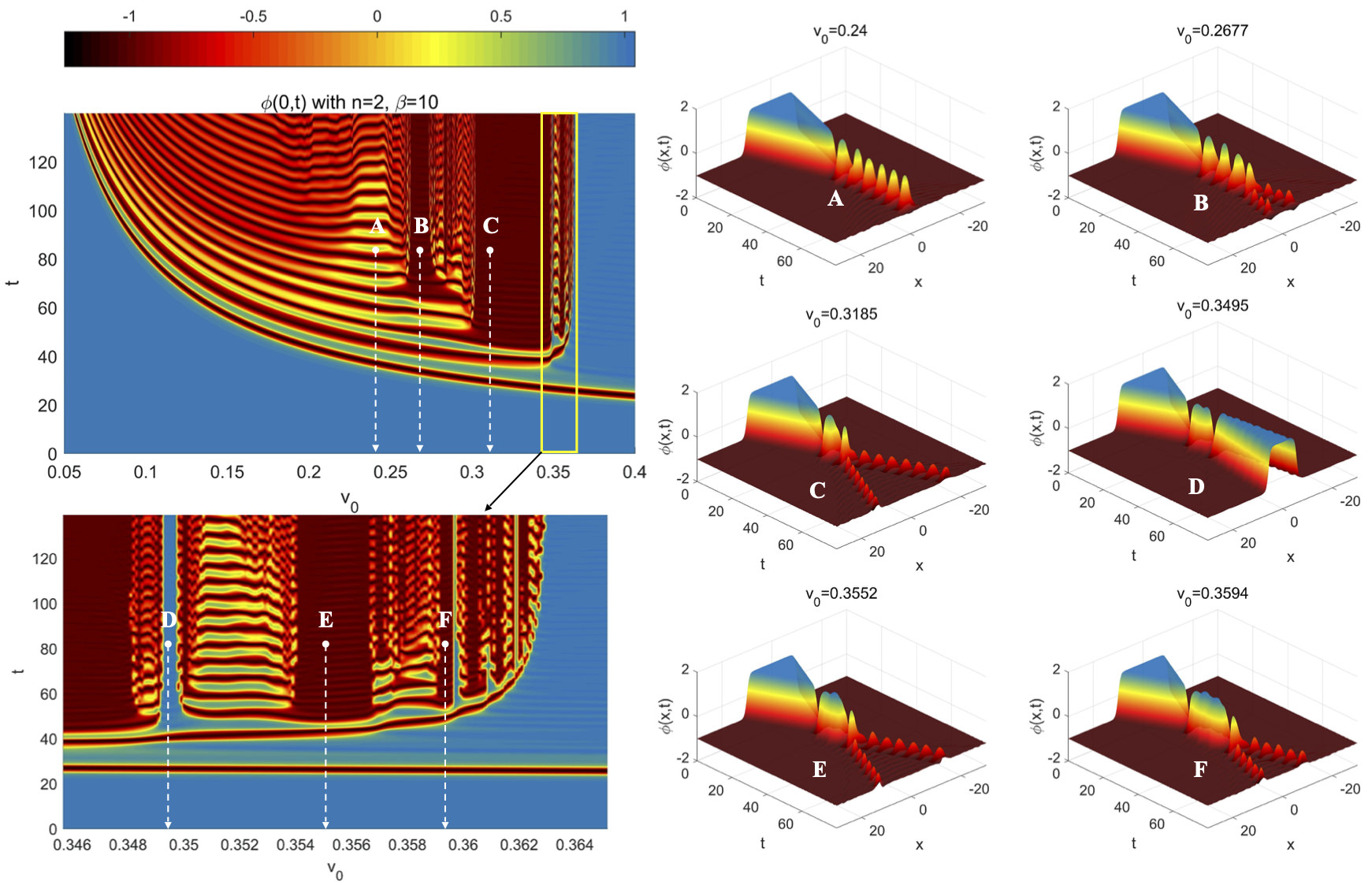}
\caption{$\phi(0,t)$ with $n=2, \beta=10$. In this case, we find two wider bion escape windows in the vicinity of $v_0=0.267$ and $v_0=0.32$. Besides, there are also three obvious two bounce windows locate around $v_0=0.3495, 0.3597$ and $0.3619$, respectively. We have chosen six representative points A-F, and plotted the corresponding field configurations. }
\label{fig_phi0t_n2_beta_10}
\end{center}
\end{figure}

In figs.~\ref{fig_n1_beta_10} and \ref{fig_phi0t_n2_beta_10}, we plot $\phi(x=0,t)$ as a function of $v_0$ for $n=1, \beta=10$ and $n=2, \beta=10$, respectively. For both cases, we can clearly see some two-bion escape windows. After magnification, narrower two-bion escape windows are found, just as higher-order bounce windows can be found by zooming in to the boundaries of any of the 2BWs. Especially, when $n=2, \beta=10$ two-bion escape windows coexist with a few 2BWs.

\begin{figure}
\begin{center}
\includegraphics[width=1\textwidth]{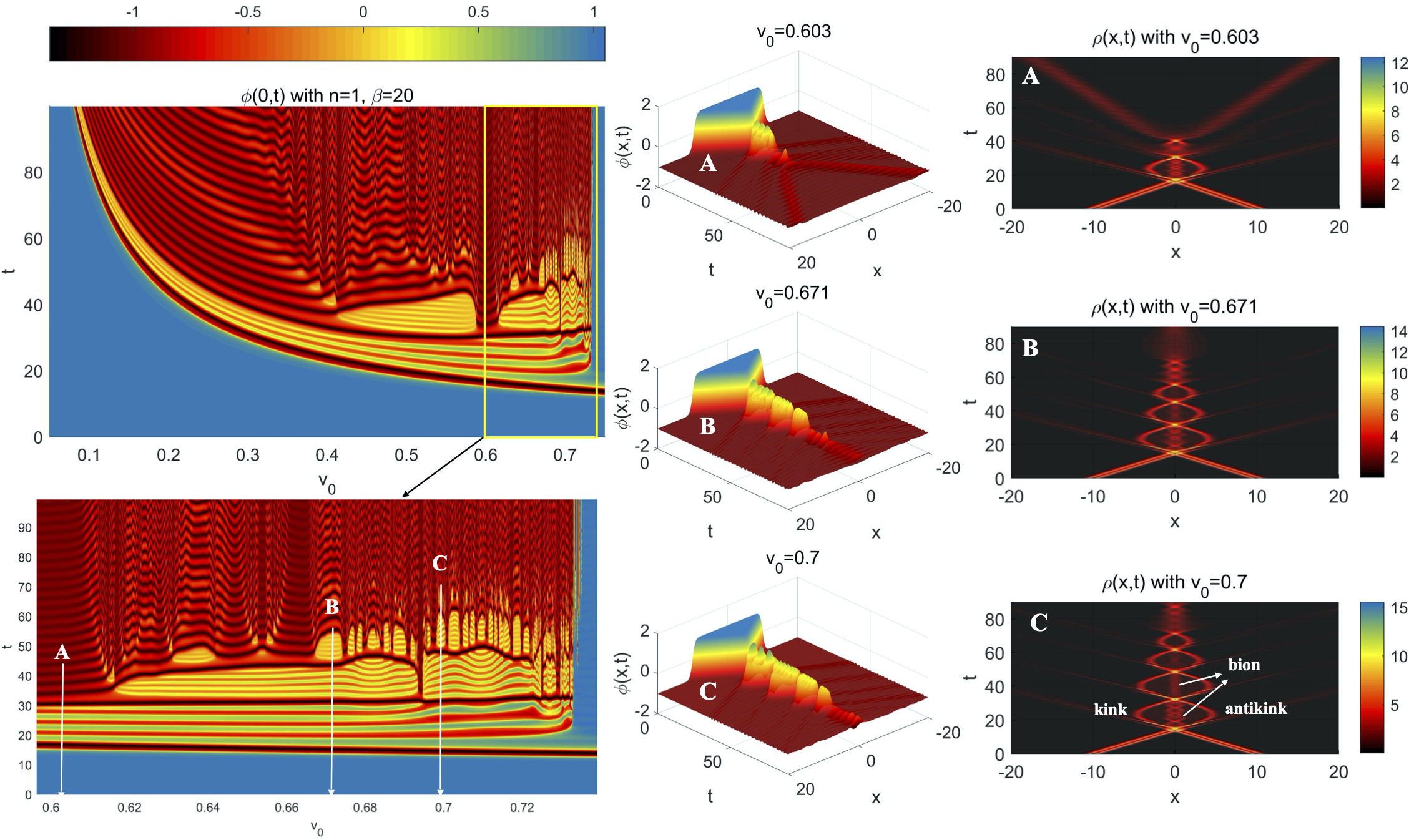}
\caption{$\phi(x=0,t)$ as a function of $v_0$ for $n=1, \beta=20$. In this case, we can see the formation of kink-bion-antikink intermediate states. There are also some two-bion escape windows, one locates around point A ($v_0=0.603$).}
\label{fig_n1_beta20}
\end{center}
\end{figure}

In addition to the two-bion escape windows, we also find some interesting intermediate states in  the case with $n=1$ and $\beta=20$.  In fig.~\ref{fig_n1_beta20}, we plotted $\phi(x=0)$ \textcolor[rgb]{0.00,0.00,1.00}{in the range $t\in[0, 140]$} and $v_0\in[0.05:0.0001:0.4]$. As can be seen from the figure, there are many yellow zones, each corresponds to a kink-bion-antikink intermediate state. Such state is constituted by a bion oscillating in the center and a kink and an antikink symmetrically moving away from the bion for a while and then come back to collide with the bion at $x=0$. Such intermediate state has also been reported in a model with double {\color{blue} kinks}~\cite{MendoncaOliveira2015}. In the range $0.1\lesssim v_0<v_c$ there is at least one such intermediate state, whose life time (the width of the lowest yellow zone) monotonically increases with $v_0$. In some narrower windows of $v_0$ one may find three or even four (see point B and point C in fig.~\ref{fig_n1_beta20}, respectively.) of such intermediate states after the collision of kinks.

 \textcolor[rgb]{0.00,0.00,1.00}{The simulations of figs.~\ref{fig_n1_beta_10}-\ref{fig_n1_beta20}  are conducted in spatial domain $x=[-50,50]$ with {\color{blue} firstly 600 and then checked with 1200 collocation points}. The tolerance option of ode45 solver is set as RelTol=$10^{-3}$ and AbsTol=$10^{-6}$. The relative error of energy is $\delta E_{\textrm{rms}} \sim 10^{-3}-10^{-4}$ for $N=600$, and $\delta E_{\textrm{rms}} \sim 10^{-5}-10^{-6}$ for $N=1200$. The representative points A, B, $\cdots$ are also checked by using more collocation points.}

\begin{figure}
\begin{center}
\includegraphics[width=1\textwidth]{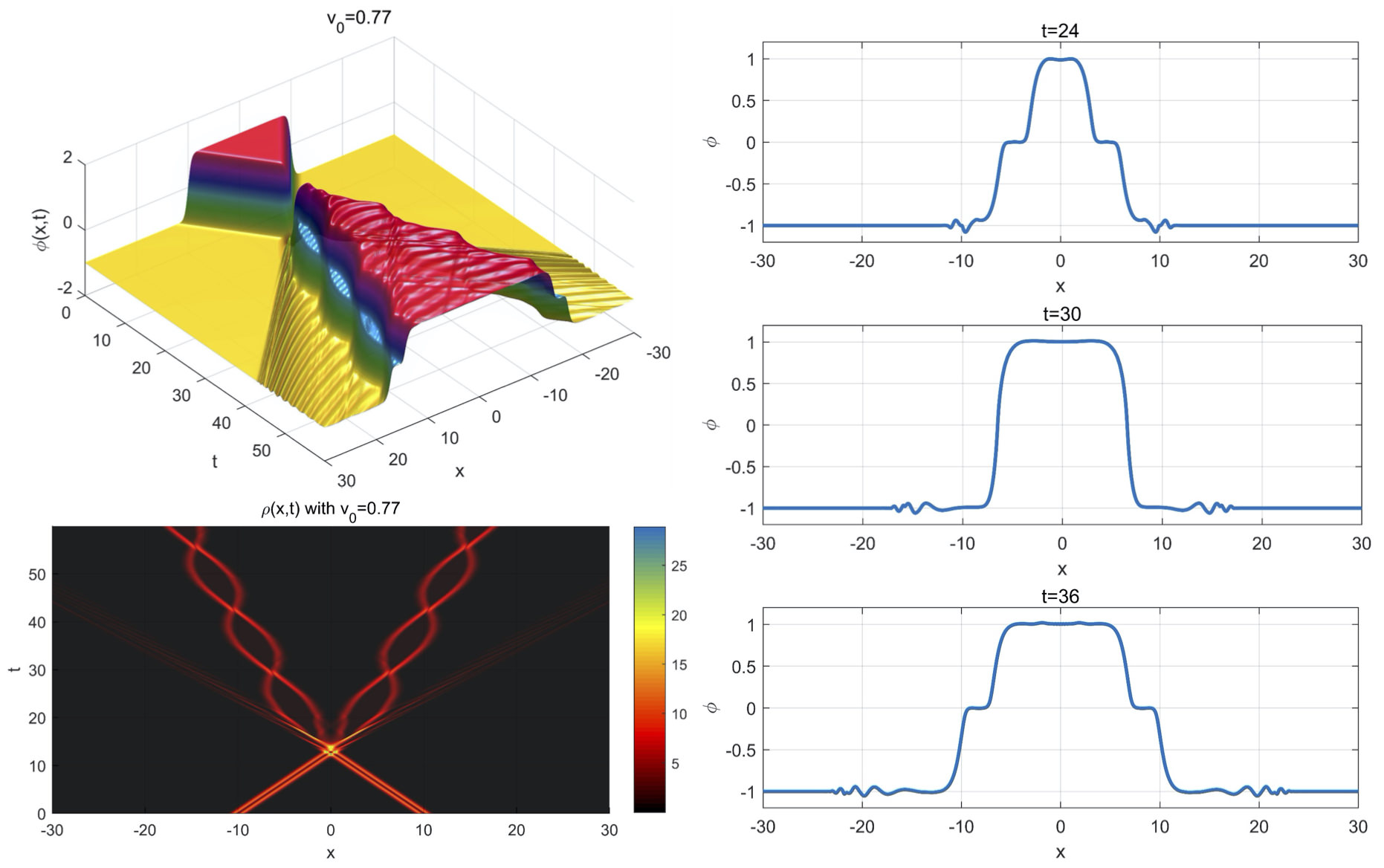}
\caption{The intertwined state found at $n=1, \beta=30$ and $v_0=0.77$. The left column {\color{blue} shows} the evolution of the scalar field configuration (upper panel) and the corresponding energy density (lower panel). The right column  {\color{blue} shows} the scalar field configurations at $t=24, 30, 36$, from which we can see that the intertwined state is a bound state of two kinks or two antikinks, which is essentially different to the bion. \textcolor[rgb]{0.00,0.00,1.00}{The simulations  are conducted in spatial domain $x=[-50, 50]$ with firstly 1500 and then checked with 3000 collocation points. Here we only display the solutions in the range $x=[-30, 30]$. The tolerance option of ode45 solver is set as RelTol=$10^{-3}$ and AbsTol=$10^{-6}$. The relative error of energy is $\delta E_{\textrm{rms}} \sim 10^{-3}$ for $N=1500$, and $\delta E_{\textrm{rms}}\sim 10^{-5}$ for $N=3000$.} }
\label{fig_n1_beta30_v077}
\end{center}
\end{figure}

The above three case studies have shown that kinks with inner structure in their energy density can have similar properties as the double kinks. Now, let us report a novel phenomenon, namely, the kink intertwined final state. This phenomenon can be observed when  $\beta$ is large enough and $v_0\gtrsim v_c$. As an example, we consider $n=1, \beta=30$ and $v_0=0.77$. The evolution of the scalar field as well as the corresponding energy density can be found in fig.~\ref{fig_n1_beta30_v077}. We can see that in this case a new structure is formed after the kink-antikink collision. This structure is similar to bion in the sense that  both of them are spatially localized oscillating solutions. The essential difference between them is that a bion is a bound state of a kink and an antikink, while the new structure we found here is a bound state of two kinks or two antikinks (see the right column of fig.~\ref{fig_n1_beta30_v077}). Another difference is that bion is formed at some initial velocities below $v_c$, but the intertwined state of kinks can be formed only when $v_0>v_c$.

\textcolor[rgb]{0.00,0.00,1.00}{We emphasize that neither the (anti-)kinks in the intermediate states nor those in the interwinded final states are the conventional ones, which connect two vacua $\phi=\pm 1$. Instead, the (anti-)kinks in these states connect only one of the vacua with the local minimum at $\phi=0$ (see fig.~\ref{fig_scalarPotential}).}

\section{Conclusion and outlook}
\label{secFour}

In this work, we investigated the kink-antikink collision in a \textcolor[rgb]{0.00,0.00,1.00}{scalar field model} with two free parameters $n$ and $\beta$. When $\beta=0$ we come back to the $\phi^4$ model, while when $\beta\gg 1$, the energy densities of the kinks (antikinks) can have rich inner structure.

Before considering the collision of kinks, we first analyzed the linear spectrum of a static kink for $n=1,2,3$ and $\beta\in[0,200]$. We found that there are at most three bound states in this range of parameters. The first bound state is the zero mode with eigenvalue $\omega_0(n, \beta)= 0$, which represents a translational mode. The second and the third bound states are two vibrational modes. As $\beta$ increases, the eigenvalue of the first vibrational mode $\omega_1(n, \beta)$ monotonically decreases, while the second vibrational mode $\omega_2(n, \beta)$ has a local minimum at $\beta_{\textrm{min}}(n)$. From fig.~\ref{fig_eingenvalues} we can see that $\beta_{\textrm{min}}(1)<\beta_{\textrm{min}}(2)<\beta_{\textrm{min}}(3)$ and $\omega_2(1,\beta_{\textrm{min}}(1))>\omega_2(2,\beta_{\textrm{min}}(2))>\omega_2(3,\beta_{\textrm{min}}(1))$.

After the analysis of the linear structure, we began to consider how the parameters $n$ and $\beta$ would {\color{blue} influence} the well-known properties of $\phi^4$ model.  We took the superposition of  a kink $\phi_K(-x_0,v_0;x,0)$ and an antikink $\phi_{\bar{K}}(x_0,-v_0;x,0)$ as the initial state, and then used the Fourier spectral method to simulate the kink-antikink collision numerically. We first calculated the critical velocity $v_c$ in the parameter scope $n=1, 2, 3$ and $\beta=[0,200]$. We found that $v_c(n, \beta)$ has a local minimum at $\beta\approx n$. When $\beta\gg1$, $v_c$ approaches to  the speed of light $c=1$.

Then we explored the impact of small $\beta$ on the width of the two bounce windows. For simplicity, we only considered the first three two bounce windows in the case with $n=1$.
We found that as $\beta$ increases, the two bounce windows first expand {\color{blue} slightly} and then {\color{blue} shrink rapidly}, and finally close at larger $\beta$. We also pointed out that although the first three two bounce windows are closed order by order with the increase of $\beta$, one cannot conclude that all the other two bounce windows are closed in this manner, as a counterexample has been found in the case with $\beta=0.9$.

After this, we began to discuss the collision phenomena in the case with large $\beta$. In this case the energy density of the kink can have more than one peak, and the kinks can have similar properties as those of the double kinks. For example, we have found many two-bion escape windows for $n=1, \beta=10$ and for $n=2, \beta=10$. In the later case we also found the coexistence of two-bion escape windows and two bounce windows. For larger value of $\beta$, for example in the case with $n=1, \beta=20$ we found the formation of some kink-bion-antikink intermediate states after the collision of kink and antikink. The number and the lifetime of these intermediate states depend on the incident velocity $v_0$. This phenomenon can also be generated by colliding two double kinks~\cite{MendoncaOliveira2015}.

Finally, we reported a novel bound state of two kinks or two antikinks. As an example, we considered the case with $n=1, \beta=30$ and $v_0=0.77$, but one can also try many other values of parameters. Two basic requirements for finding this phenomenon are $\beta\gg 1$ and $v_0>v_c$.

This work revels the fact that kinks with inner structure in their energy density may have similar properties as those of the double kink solutions. Both can have two-bion escape final states and kink-bion-antikink intermediate states after a collision. When $v_0>v_c$ we found a new spatially localized oscillating structure, which to our knowledge, has not been reported before. Unlike the bion, which is a bound state of a kink and an antikink, the new structure we found here is a bound state between two kinks or two antikinks.

As an outlook, we would like to point out that we have not cover all the parameter ranges, for example, the cases with $n>3$  are not discussed. Even for $n=1, 2, 3$ we cannot claim that  we have found all the distinct phenomena. As we have shown in subsection~\ref{secCollision}, the collision result sensitively depends on the values of $\beta$ and $v_0$, but we have only studied a few representative values of $\beta$. Therefore, it would be possible to find other new phenomena by considering different parameter settings {\color{blue} from} ours. Besides, the superpotential we taken in eq.~\eqref{eqSuper} leads to a $\phi^4$ type of kink solution, it is easy to generate other kink (for example a sine-Gorden type of kink) or double kink solutions by simply taking different superpotentials. As a future direction, one can consider the collision of these kinks in our model. If one would like to go beyond the present model, there are many other noncanonical kink models such as those studied in refs.~\cite{ZhongLiu2014,ZhongGuoFuLiu2018}. At present time, only a few works considered the interactions of noncanonical kinks \cite{GomesMenezesNobregaSimas2014}, so this field is worth further investigation.
It is also interesting, despite challenging, to understand how the intermediate and final states we reported above are formed.
Finally, it would be interesting to discuss the application of the intertwined two kink final states as a cosmological reheating mechanism, in parallel to the previous work \cite{TakamizuMaeda2004}.

\section*{Acknowledgment}
This work was supported by the National Natural Science Foundation of
China (Grant Numbers 11847211, 11605127, 11875151, 11522541, 11405121, and 11375075), the Fundamental Research Funds for the Central Universities (Grant No. xzy012019052), and by China Postdoctoral Science Foundation (Grant No. 2016M592770).

\providecommand{\href}[2]{#2}\begingroup\raggedright\endgroup

\end{document}